%% file: manuscript_icassp.tex
\documentclass{article}
\usepackage{spconf,amsmath,graphicx}
\usepackage[T1]{fontenc}
\usepackage[utf8]{inputenc}
\usepackage{epsfig}
\usepackage{amsmath}
\usepackage{amssymb}
\usepackage{cite}
\usepackage{color}
\usepackage{url}
\usepackage{amsthm}
\usepackage{algorithm}
\usepackage{algorithmic}
\usepackage{multirow}
\usepackage{subfigure}
\usepackage{tabularx}
\usepackage{array}
\usepackage{varwidth}
\usepackage{setspace}
\usepackage{comment}
\usepackage{mathtools, cuted}
\usepackage{float}

\title{General Total Variation Regularized Sparse Bayesian Learning for Robust Block-Sparse Signal Recovery}

\name{Aditya~Sant$^{*}$, Markus~Leinonen\textsuperscript{\textdagger} and Bhaskar~D.~Rao$^{*}$\thanks{The work of A. Sant and B. D. Rao has been financially supported by ONR Grant No. N00014-18-1-2038 and the UCSD Center for Wireless Communications. The work of M. Leinonen has been financially supported in part by Walter Ahlström Foundation through Tutkijat Maailmalle program, Infotech Oulu, the Academy of Finland (grant 323698) and (grant 319485), and Academy of Finland 6Genesis Flagship (grant 318927).}}
\address{*Department of Electrical and Computer Engineering, University of California San Diego \\
\textsuperscript{\textdagger}Centre for Wireless Communications -- Radio Technologies, University of Oulu, Finland}

\makeatletter
\newcommand{\multiline}[1]{%
\begin{tabularx}{\dimexpr\linewidth-\ALG@thistlm}[t]{@{}X@{}}
#1
\end{tabularx}
}
\makeatother

\graphicspath{{Figure/}}
\DeclareGraphicsExtensions{.pdf}
\newcommand{\bibdir}{Bib/}
\newcommand{\mybibliography}{\bibliography{\bibdir/jour_short,\bibdir/conf_short,\bibdir/refs}}
\input{Commands}

\begin{document}
\ninept
\sloppy 
\maketitle
\begin{abstract}
Block-sparse signal recovery without knowledge of block sizes and boundaries, such as those encountered in multi-antenna mmWave channel models, is a hard problem for compressed sensing (CS) algorithms. We propose a novel Sparse Bayesian Learning (SBL) method for block-sparse recovery based on popular CS based regularizers with the function input variable related to total variation (TV). Contrary to conventional approaches that impose the regularization on the signal components, we regularize the SBL hyperparameters. This iterative TV-regularized SBL algorithm employs a majorization-minimization approach and reduces each iteration to a convex optimization problem, enabling a flexible choice of numerical solvers. The numerical results illustrate that the TV-regularized SBL algorithm is robust to the nature of the block structure and able to recover signals with both block-patterned and isolated components, proving useful for various signal recovery systems. 
%
\end{abstract}


\section{Introduction}
Block-sparse signal recovery has various applications in wireless communication, audio, and image processing. We are primarily interested in such signal recovery for mmWave channel estimation where the received signal is composed of angular multipath components that impinge on the receive antenna as clustered rays \cite{chan_model1,chan_model2}. One of the main challenges with block-sparse recovery is to model inter-element dependency, in addition to the sparsity constraint. Since block sizes can be unequal and block boundaries are unknown, the number of possible signal ``blocks'' involved in the search grows exponentially with the grid size. Hence, there is a need to impose structure on signal recovery algorithms. This also presents an inherent trade-off between computational complexity and block-sparse modeling for arbitrarily sized blocks with unknown boundaries.
%
%

Considering known block partitions, compressed sensing (CS) \cite{Candes-Romberg-Tao-06,Donoho-06} algorithms have been modified for block-sparse signals, which include Group-Lasso \cite{group_lasso}, Group Basis Pursuit \cite{group_basis_pursuit}, Model-based CoSaMP \cite{model_cosamp}, and Block-OMP \cite{block_omp}. Early attempts for block-sparse recovery under unknown block partitions include Struct-OMP in \cite{struct_omp} and the method based on graphical models in \cite{bm_paper}.
The works \cite{bsbl_main,bsbl_algorithms} were the first SBL approaches and developed the Block SBL (BSBL) algorithm. Using Bayesian CS, \cite{spike_slab_prior} incorporated a spike-and-slab prior to model both block and individual sparsity. 

The Pattern-coupled SBL (PC-SBL) \cite{pc_sbl} enforced block-sparse structures by coupling the underlying SBL parameters. The Non-uniform Burst Sparsity algorithm proposed in \cite{burst_sparsity_learning} improved on \cite{pc_sbl} through Variational Bayesian Inference. Coupled priors were also used in the Extended-BSBL (EBSBL) method in \cite{alt_ext_sbl}. The works in \cite{clustered_sparsity,clustered_sparsity_2} enforce block structures using a cluster-structured prior.

We propose a novel total variation (TV) based regularizer for SBL to promote block-sparse signal recovery. Specifically, we enforce block sparsity in the hyperparameter space of SBL to promote uninterrupted zero regions of the estimated signal. This is achieved using common regularizers from CS acting on the hyperparameter TV input variable, instead of signal components.
%
%
%
The framework is quite general and allows for an exploration of a wide range of regularizers utilizing the experience from CS. To the best of our knowledge, this is the first work to apply a TV type penalty in the hyperparameter space of SBL to encourage block-sparsity. Majorization-minimization is used to convexify the proposed SBL formulation and develop an iterative algorithm. Numerical results show that by inducing a soft TV prior on the parameters, the TV-regularized SBL method is robust to sparsity structure; the algorithm attains definitive recovery from strict block-sparsity to fully random sparsity.  

\section{New Total Variation Regularizers for Block-Sparse Signal Recovery via SBL}\label{sec_SBL}
We consider a multiple measurement vector (MMV) problem which involves simultaneous estimation of $L$ block-sparse source vectors ${\xb_{l}\in\Cbb^{N}}$ from a collection of noisy linear measurements.
\begin{equation}\label{eq:measurements}
\yb_{l}=\Ab\xb_{l}+\nb_{l},~l=1,\ldots,L,
\end{equation}
where ${\yb_{l}\in\Cbb^{M}}$ is a measurement vector at time instant $l$, $\Ab\in\Cbb^{M\times{N}}$ is a fixed known measurement matrix, and ${\nb_{l}\sim\mathcal{C}\mathcal{N}(\zerob,\lambda\Ib)}$ is a noise vector, independent of $\xb_{l}$. Source vectors and noise vectors are assumed to be independent and identically distributed (i.i.d.) across the time instants. The same sparsity pattern is shared among the collection of vectors $\{\xb_{l}\}_{l=1}^{L}$.
Thus, the signal ensemble $\Xb=[\xb_{1}\cdots\xb_{L}]$ is \emph{block-row-sparse}. We assume that both the block sizes and their locations are \emph{unknown}.
 
The early works imposed a block structure on SBL inference through a specific deterministic or stochastic parameterization of the signal; for example, BSBL \cite{bsbl_algorithms} relies on a pre-determined block partition. Whereas, to handle dynamic block sizes, algorithms like PC-SBL \cite{pc_sbl,burst_sparsity_learning} impose \emph{explicit coupling} on the variables. Our approach improves on one limitation of such coupling based approaches: reduced sensitivity to isolated spurious components. We first provide a brief overview of the SBL framework for sparse signal recovery. 
%
%
\subsection{SBL Framework: Generalized Cost Function} \label{sec_sbl_overview}
There are various advantages of SBL for MMV sparse signal recovery, motivating our choice: \textit{(i)} The M-SBL  \cite{sbl_wipf_mmv} parameter estimation abstracts each row of $\Xb$ by a single parameter ($\gamma_{i}$), reducing the number of parameters to be estimated from $NL$ to $N$ compared to CS approaches; \textit{(ii)} It falls under the class of methods that are correlation-aware which have shown superior ability to find sparse solutions \cite{pal_ppv_sbl};
\textit{(iii)} SBL shows great promise for sparse signal recovery under correlated sources and ill-conditioned dictionaries \cite{rohan_sbl_correlated}. 
%

We now describe the SBL inference. 
With an additive Gaussian noise model \eqref{eq:measurements}, the SBL framework \cite{sbl_wipf_mmv} 
assumes a parametric Gaussian distribution for each signal ${\xb_{l}\in\Cbb^{N}}\ (l\in\{1,\dots,L\})$ as 
$p(\xb_{l};\gammab)=\Ccal\Ncal(\zerob,\Gammab) = \frac{1}{\sqrt{(2\pi)^{N}|\Gammab|}}\exp\left(-\frac{1}{2}\xb_{l}\herm\Gammab\inv\xb_{l}\right)$,
where ${\gammab=[\gamma_{1}\cdots\gamma_{N}]\tran\in\Rbb_{+}^{N}}$ is a vector of \emph{hyperparameters}, adjusting the variance of each signal component $x_{l,i}$, ${i=1,\ldots,N}$, and ${\Gammab\triangleq\diag(\gammab)}$. The hyperparameter values $\gammab$ reflect the sparsity profile of the block-row-sparse $\Xb$; a suitable prior on $\gammab$ can lead $\xb_{l}$ to model many interesting sparse priors, e.g., Gaussian scale mixtures.

The posterior density $p(\xb_{l}|\yb_{l};\gammab)$  
is also Gaussian as $\Ccal\Ncal(\mub_{\xb_{l}|\yb_{l};\gammab},\Sigmab_{\xb|\yb;\gammab})$, where
\begin{equation}\label{eq:conditional_mean_covariance}
    \mub_{\xb_{l}|\yb_{l};\gammab}=\lambda\inv\Sigmab_{\xb|\yb;\gammab}\Ab\herm\yb_{l},\  \Sigmab_{\xb|\yb;\gammab}={\big(\lambda\inv\Ab\herm\Ab+\Gammab\inv\big)}\inv.
\end{equation} 
For a given ${\gammab}$, the estimate of each signal $\{\xb_{l}\}_{l=1}^{L}$ is formed as  ${\hat{\xb}_{l,\mathrm{SBL}}=\mub_{\xb_{l}|\yb_{l};\gammab}}$ according to \eqref{eq:conditional_mean_covariance}. Following \cite{sbl_wipf_mmv}, the hyperparameter estimation is done through Type-\MakeUppercase{\romannumeral 2} maximum a posteriori (MAP) estimation of the posterior $p(\gammab|\yb_{1},\ldots,\yb_{L})$ over $\gammab$, i.e.,  
\begin{equation}\label{eq:posterior_gamma_opt}
\gammab^*=\disp\underset{\gammab\succeq\zerob}{\argmin}~L\,\log\,|\Sigmab_{\yb}|+\textstyle\sum_{l=1}^{L}\yb_{l}\herm\Sigmab_{\yb}^{-1}\yb_{l}-\log\,p(\gammab),
\end{equation}
where ${\Sigmab_{\mathbf{y}}=\lambda\Ib+\Ab\mathbf{\Gamma}\Ab\herm}$ is the measurement covariance matrix and $\log\,p(\gammab)$ is the \emph{hyperprior} on $\gammab$.
The expression in \eqref{eq:posterior_gamma_opt} is the generalized MMV SBL cost function. This optimization is \emph{non-convex} due to the concave term $\log\,|\Sigmab_{\yb}|$; convexity of $\log\,p(\gammab)$ depends on the prior. We elaborate on the optimization strategies in Sec. \ref{sec_optimization}.
\subsection{SBL with Novel TV-based Regularizers}\label{sec_novel_tv_regularizer}
%
%
Most existing approaches enforce structure by working directly on $\xb_l$, a more challenging and less efficient approach for the complex block-sparsity problem. \emph{Informative} priors/regularizers $\log\,p(\gammab)$ in \eqref{eq:posterior_gamma_opt} can help improve inference \cite{sbl_laplace_prior,wipf_sparse_priors,tharun_rao}.
Although block-sparse methods have been developed using SBL, the priors are often strong, thereby biasing the methods and making them brittle.
Surprisingly, some simple regularizers seem to have been overlooked and we show them to be quite effective and, more importantly, robust.

To support block-sparse solutions, we opt for a regularizer that combines various sparse regularizers developed in CS, with Total Variation (TV) \cite{tv_osher,tv_rudin,tv_vogel}. 
To this end, we denote the hyperprior as $\beta\,T(\gammab)\triangleq-\log\,p(\gammab)$, where $\beta$ is a non-negative weighting parameter and $T(\cdot)$ is a general TV-type penalty of vector $\gammab$. Regardless of $T(\cdot)$, we refer to our developed method collectively as \textbf{TV-SBL}.

We now describe the motivation of our regularizer $T(\gammab)$. For maximally (random) sparse solutions, an appropriate choice is $T(\gammab)=\sum_{i} I(\gamma_{i})$, where the indicator function $I(\cdot)$ is an exact measure of sparsity as $I(\gamma_{i})=1$ for $\gamma_{i}>0$ and zero otherwise. Since the function $T(\gammab)$ is intractable, many surrogate measures have been used, the most common one being the $\ell_1$-norm in CS. 

Using the indicator function to help block-sparsity, the main driver of our work is 
$T(\gammab) = \sum_{i} I(|\gamma_{i} - \gamma_{i-1}|)$,
i.e., TV on $\gammab$. This assumes equal variances of the entries within a block and thus optimally counts the number of edges in the underlying signal. Armed with this ideal measure, we can use tractable measures developed in CS on the TV inspired input variable
$|\gamma_{i} - \gamma_{i-1}|$ to identify appropriate block structures. It is noteworthy that imposing this regularizer on the hyperparameters rather than the source vectors $\xb_l$ is an important distinction and also key to the success of our approach. CS theory has developed many regularizers which are monotonically increasing and concave on the positive orthant to promote sparsity. We discuss two options to illustrate the potential of the TV framework.

\noindent \textit{1) Linear TV: Conventional Smoother}

\noindent The Linear TV regularizer is equivalent to the $\ell_1$ penalty in CS and is given by the form 
\begin{equation}\label{eq_sbl_tv_case1}
T(\gammab) = \textstyle\sum_{i=2}^{N}|\gamma_{i}-\gamma_{i-1}|.
\end{equation}
It has also been used in different signal processing applications
to preserve edges and enforce local smoothness. We use this convex regularizer to enforce a block structure in the recovered signal. In addition to the signal regions, this penalty is found to denoise the zeros more effectively than the unregularized SBL algorithm. 

\noindent \textit{2) Log TV: CS-based Regularizer}

\noindent Another widely used regularizer in CS is $\sum_{i=1}^{N}\log(|x_i|+\epsilon),$ where $\epsilon$ is a positive stability parameter. This regularizer employs an iterative reweighted $\ell_{1}$ minimization algorithm and has been shown to yield superior recovery \cite{shen_log_sum,candes_boyd_log_sum}.
Utilizing this regularizer for block-sparsity, the Log TV regularizer is given by
\begin{equation}\label{eq_sbl_tv_case2}
T(\gammab) = \textstyle\sum_{i=2}^{N}\mathrm{log}(|\gamma_{i}-\gamma_{i-1}|+\epsilon).
\end{equation}
As in the CS, the Log TV based approach is found to be more effective than the Linear TV. This is due to its better resemblance to $\ell_{0}$-norm \cite{candes_boyd_log_sum}, allowing more signal variance differences within a block and restraining small (faulty) signal estimate components to emerge. 

\section{Optimization Approaches for TV-SBL}\label{sec_optimization}
There are many options for minimizing the general SBL objective function. We apply the majorization-minimization (MM) approach and derive an iterative algorithm for minimizing the TV-SBL cost. 

\subsection{Optimization of TV-SBL with Linear TV} 
\noindent The TV-SBL optimization \eqref{eq:posterior_gamma_opt} for the Linear TV regularizer in \eqref{eq_sbl_tv_case1} is
\begin{equation}\label{eq_sbl_tv_cost_function_linear}
    \gammab^*\!=\!\disp\underset{\gammab\succeq\zerob}{\argmin}\,L\,\log|\Sigmab_{\yb}|+\textstyle\sum_{l=1}^{L}\yb_{l}\herm\Sigmab_{\yb}^{-1}\yb_{l}+\beta\,\sum_{i=2}^{N}\!|\gamma_{i}-\gamma_{i-1}|.
\end{equation}
Using the MM technique similar to \cite{wipf_sbl_mm}, we majorize the concave term $\log|\Sigmab_{\yb}|$ and solve iteratively a sequence of convex optimization problems. We majorize (i.e., linearize) $\log|\Sigmab_{\yb}|$ by its first-order Taylor approximation at point $\Gammab^{(j)}$, i.e.,
\begin{equation}\label{eq_sbl_tv_majorize_log_det}
    \begin{array}{ll}
         \log|\lambda\Ib+\Ab\Gammab\Ab\herm| \le  \log|\lambda\Ib+\Ab\Gammab^{(j)}\Ab\herm| + \\ \trace\big((\Sigmab_{\yb}^{(j)})\inv\Ab\Ab\herm[\Gammab-\Gammab^{(j)}]\big),
    \end{array}
\end{equation}
where the superscript $j$ denotes the MM iteration index. Using \eqref{eq_sbl_tv_majorize_log_det}, at iteration $j$, we end up with solving the convex problem  
\begin{equation}\label{eq_sbl_tv_cost_function_linear_mm}
    \begin{array}{ll}
    \gammab^{(j+1)}=&\hspace{-3mm}\disp\underset{\gammab\succeq{\zerob}}{\argmin}~L\,\trace\left(\big(\Sigmab_{\yb}^{(j)}\big)\inv\Ab\Gammab\Ab\herm\right) \\
    &\hspace{-3mm}+\textstyle\sum_{l=1}^{L}\yb_{l}\herm\Sigmab_{\yb}^{-1}\yb_{l} + \beta\sum_{i=2}^{N}|\gamma_{i}-\gamma_{i-1}|,
    \end{array}
\end{equation}
and then updating $\Sigmab_{\yb}^{(j)}$ using the newly obtained $\gammab^{(j+1)}$.

\subsection{Optimization of TV-SBL with Log TV}

\noindent The TV-SBL optimization \eqref{eq:posterior_gamma_opt} for the Log TV regularizer in \eqref{eq_sbl_tv_case2} is 
\begin{equation}\label{eq_sbl_tv_cost_function_log}
    \begin{array}{ll}
    \gammab^*=&\hspace{-3mm}\disp\underset{\gammab\succeq\zerob}{\argmin}~L\,\log|\Sigmab_{\yb}|+\textstyle\sum_{l=1}^{L}\yb_{l}\herm\Sigmab_{\yb}^{-1}\yb_{l}\\
    &\hspace{-3mm}+\beta\,\textstyle\sum_{i=2}^{N}\mathrm{log}(|\gamma_{i}-\gamma_{i-1}|+\epsilon).
    \end{array}
\end{equation}
Similar to the Linear TV case above, we apply the MM approach for \eqref{eq_sbl_tv_cost_function_log}. Besides majorizing the $\log|\Sigmab_{\yb}|$ term via \eqref{eq_sbl_tv_majorize_log_det}, we majorize the concave Log TV penalty \eqref{eq_sbl_tv_case2} by its first-order Taylor approximation at points $(\gamma_{i}^{(j)}-\gamma_{i-1}^{(j)})$, $i=2,\ldots,N$, i.e.,
\begin{equation}\label{eq_sbl_tv_majorize_log_tv}
    \begin{array}{ll}
         \mathrm{log}(|\gamma_{i}-\gamma_{i-1}|+\epsilon) \le  \mathrm{log}(|\gamma_{i}^{(j)}-\gamma_{i-1}^{(j)}|+\epsilon) +   \frac{|\gamma_{i}-\gamma_{i-1}|}{|\gamma_{i}^{(j)}-\gamma_{i-1}^{(j)}|+\epsilon}. 
    \end{array}
\end{equation}
Thus, at iteration $j$, we solve the convex problem 
\begin{equation}\label{eq_sbl_tv_cost_function_log_mm}
    \begin{array}{ll}
    \gammab^{(j+1)}\!=\,\disp\underset{\gammab\succeq{\zerob}}{\argmin}~L\,\trace\left(\big(\Sigmab_{\yb}^{(j)}\big)\inv\Ab\Gammab\Ab\herm\right)\\ 
    + \textstyle\sum_{l=1}^{L}\yb_{l}\herm\Sigmab_{\yb}^{-1}\yb_{l}
    +\beta\textstyle\sum_{i=2}^{N}\frac{1}{|\gamma_{i}^{(j)}-\gamma_{i-1}^{(j)}|+\epsilon}\,|\gamma_{i}-\gamma_{i-1}|, 
    \end{array}
\end{equation}
followed by updating $\Sigmab_{\yb}^{(j)}$ using the newly obtained $\gammab^{(j+1)}$.

\subsection{Convex Solver Implementation of TV-SBL}
Any convex optimization package can be implemented to solve \eqref{eq_sbl_tv_cost_function_linear_mm} and \eqref{eq_sbl_tv_cost_function_log_mm}. Algorithm \ref{algo_conv_opt_full} presents the implementation of TV-SBL via the widely used CVX optimization package \cite{cvx} to facilitate easy adoption and experimentation.
One key step is to handle the matrix inverse in $\yb_{l}\herm\Sigmab_{\yb}^{-1}\yb_{l}$ through the Schur's complement equivalence \cite[Appendix A5.5]{boyd_convex} by introducing the Hermitian symmetric matrix variables $\Zb_{\mathrm{cvx},l}\in\mathbb{S}^{M\times M}$, $l=1,\ldots,L$.
%
%
%
\begin{algorithm}[H]
    \caption{CVX Solver for TV-SBL}
    \textbf{Input:} $\Ab, \Yb, \gammab^{(0)}$, $\lambda$, $\beta$, $\epsilon$, $j_{\mathrm{max}}$ \\
    \textbf{Output:} $\mub_{\xb_{l}|\yb_{l};\gammab}$ ($\forall \ l\in\{1,\dots,L\}$)
    \begin{algorithmic}[1]
        \FOR{$j=0$ to $j_{\mathrm{max}}-1$}
            \STATE Evaluate $[{\Sigmab_{\mathbf{y}}^{(j)}]^{-1}=(\lambda\Ib+\Ab\mathbf{\Gamma}^{(j)}\Ab\herm})^{-1}$
            \STATE CVX variables: ${\gammab_{\mathrm{cvx}}\in\mathbb{R}^{N}}$, ${\Zb_{\mathrm{cvx},l}\in\mathbb{S}^{M\times M}}$, $l=1,\dots,L$ \\
                \STATE  \textbf{minimize}: 
                $L\,\trace\big{[}(\Sigmab_{\yb}^{(j)}\big)\inv\Ab\Gammab_{\mathrm{cvx}}\Ab\herm\big{]}+\sum_{l=1}^{L}\trace(\Zb_{\mathrm{cvx},l})+\beta \ T(\gammab_{\mathrm{cvx}})$ \hfill\COMMENT{$T(\gammab_{\mathrm{cvx}})\rightarrow$ Linear TV \eqref{eq_sbl_tv_case1} or Log TV \eqref{eq_sbl_tv_case2}} 
                \STATE \textbf{subject to}: 
                $\gammab_{\mathrm{cvx}}\succeq{\zerob}$, \ 
                $\begin{bmatrix}
                \Zb_{\mathrm{cvx},l} & (\yb_{l}\yb_{l}\herm)^{1/2} \\
                (\yb_{l}\yb_{l}\herm)^{1/2} & \lambda\Ib+\Ab\mathbf{\Gamma_{\mathrm{cvx}}}\Ab\herm
                \end{bmatrix}\succeq{\zerob}$ ($\forall \ l\in\{1,\dots,L\}$)
            \STATE $\gammab^{(j+1)}\leftarrow\gammab_{\mathrm{cvx}}$
        \ENDFOR
        \STATE Evaluate $\mub_{\xb_{l}|\yb_{l};\gammab}$ ($\forall \ l\in\{1,\dots,L\}$) using $\gammab^{(j_{\mathrm{max}})}$ in \eqref{eq:conditional_mean_covariance}
    \end{algorithmic}
    \label{algo_conv_opt_full}
\end{algorithm}

\section{Numerical Results}\label{sec:results}

This section provides the numerical results for the CVX solver described in Algorithm \ref{algo_conv_opt_full}.
For the MMV setup \eqref{eq:measurements}, we consider a signal length of $N=150$ with $M=20$ measurements and $L=5$ snapshots. We form the dictionary $\mathbf{A}\in \mathbb{R}^{M\times N}$ by first drawing its elements from a Gaussian distribution, and then normalizing the columns as $\| \cdot \|_2=1$. The signal ensemble $\Xb$ contains $K=10$ non-zero rows and each non-zero element is drawn from $\Ncal(0,1/K)$. We consider three classes of block-sparse signals:

\noindent \textit{(i)} \textit{Homogeneous} block-sparse signal with $2$ blocks of length $5$ each; 

\noindent\textit{(ii)} \textit{Random} sparse signal with $10$ randomly placed components (which are thus mostly isolated);

\noindent\textit{(iii)} \textit{Hybrid} sparse signal with $1$ block of length $4$, $1$ block of length $3$, and $3$ isolated components.

Each noise signal $\nb_{l}$, $l=1,\ldots,L$, is generated from $\Ncal(0,\sigma_{n}^{2})$ with variance $\sigma_{n}^{2}$ chosen so that the Signal-to-noise ratio (SNR), 
%
$
   10  \mathrm{log}_{10}\left(\frac{\mathbb{E}\ ||\Ab\xb_{l}||^{2}}{\mathbb{E} \ ||\nb_{l}||^{2}}\right)$\footnote{All expectations $\mathbb{E}[\cdot]$ are evaluated over 200 Monte Carlo trials.},
%
varies from $0$ to $20$ dB. 
%

\begin{figure}[t!]
\centering
\subfigure[]{\includegraphics[width=0.25\textwidth]{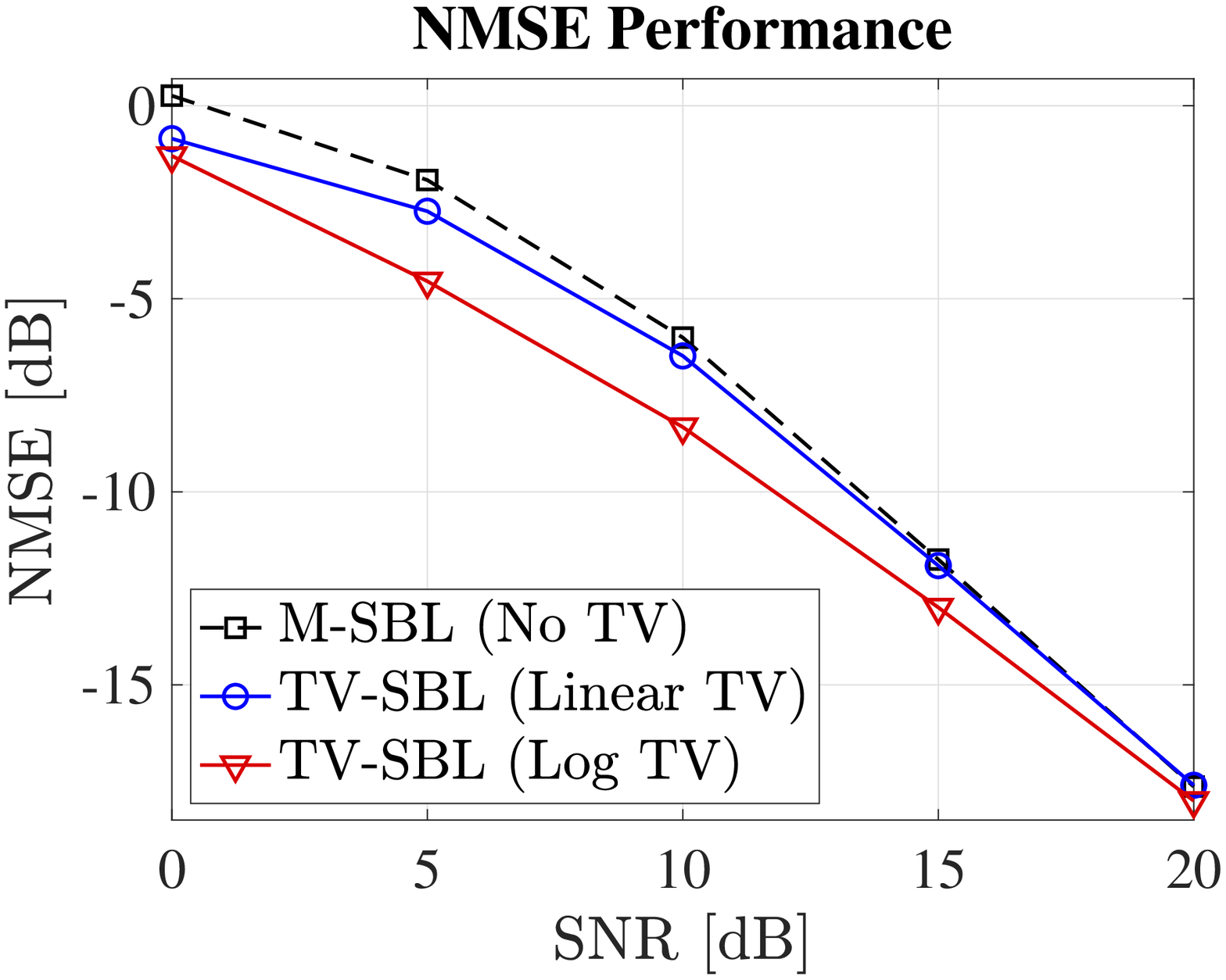}}\hspace{-4mm}
\subfigure[]{\includegraphics[width=0.25\textwidth]{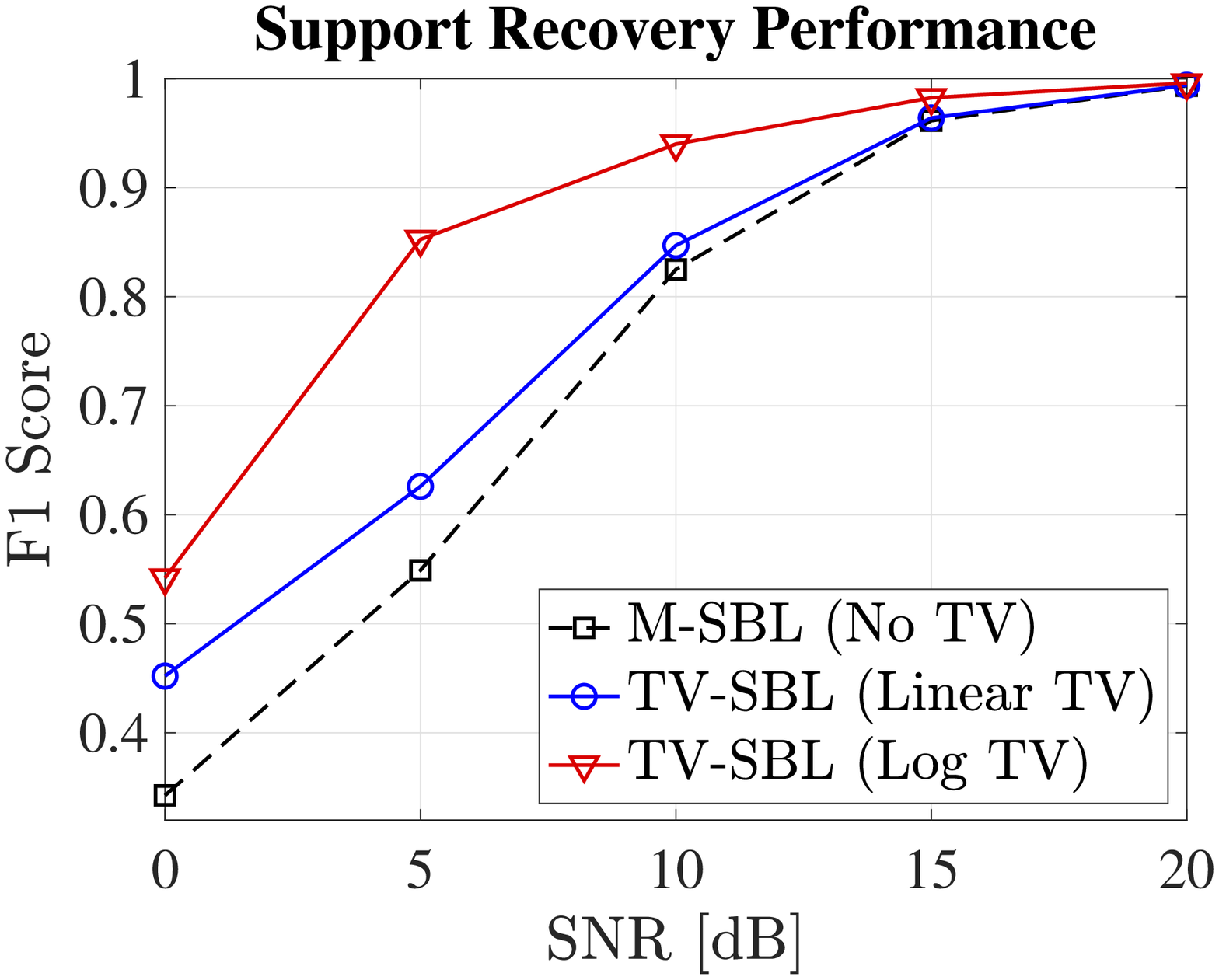}}\vspace{-5mm}
\caption{Comparison of Linear TV and Log TV penalties under homogeneous block-sparsity (see Fig.~\ref{fig_perf_comparison_random_M_20_N_150_blocks_5_2}(a)): (a) NMSE and (b) F$_1$-Score.}\vspace{-4mm}
\label{fig_tv_compare_random_M_20_N_150_blocks_5_2}
\end{figure}

\begin{figure*}[t!]
\subfigure[]{\includegraphics[width=0.28\textwidth]{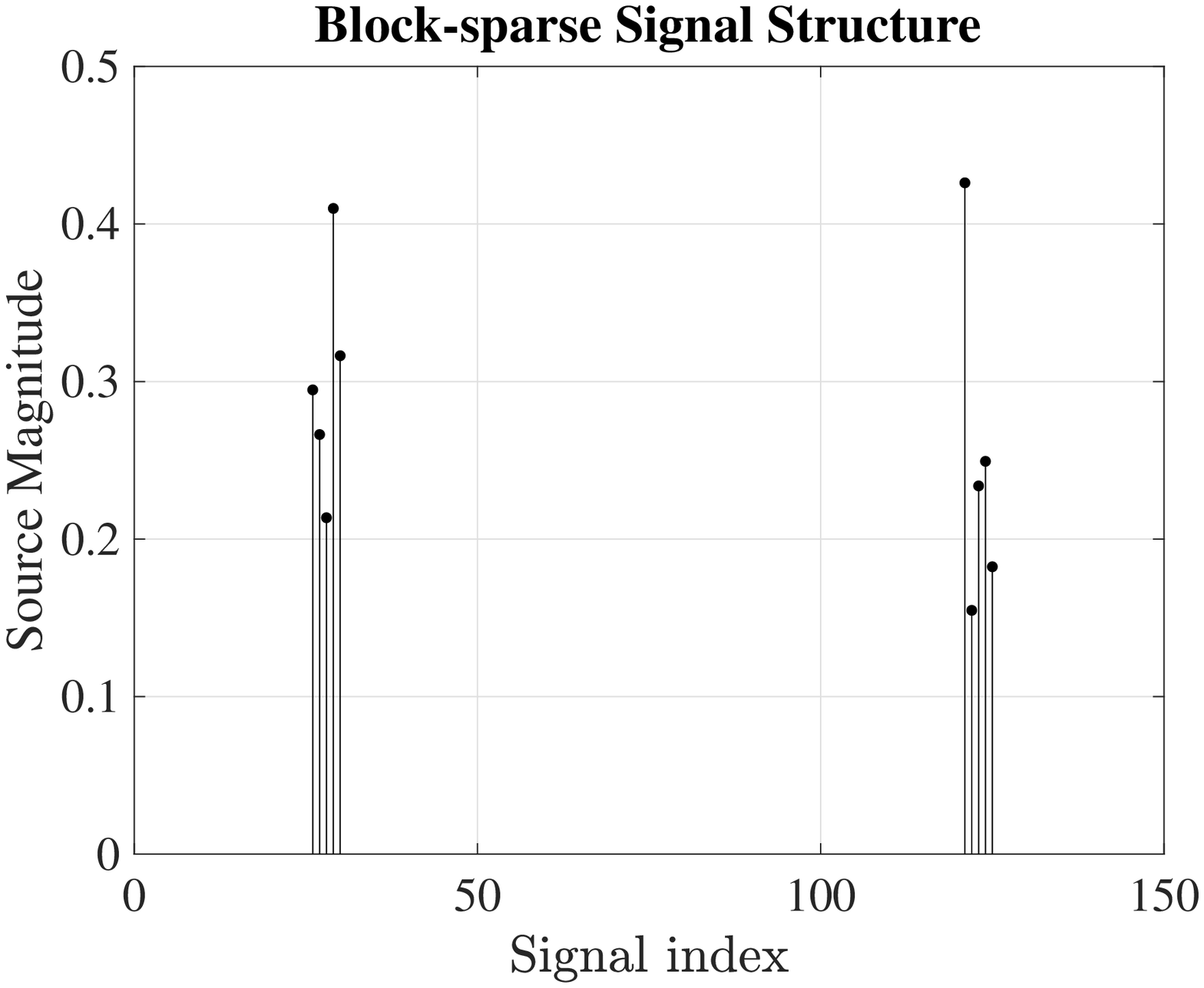}} 
\subfigure[]{\includegraphics[width=0.33\textwidth]{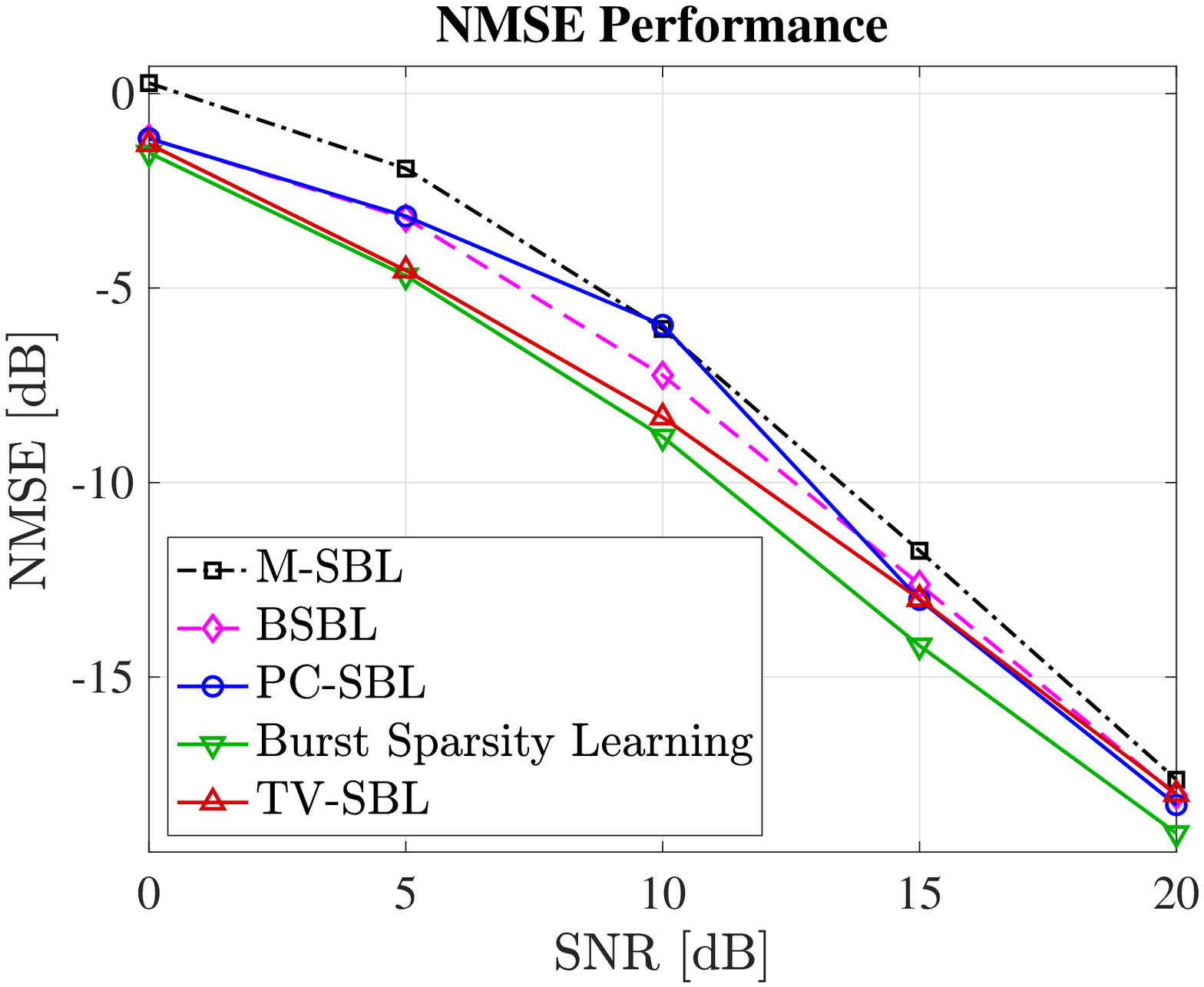}} 
\subfigure[]{\includegraphics[width=0.33\textwidth]{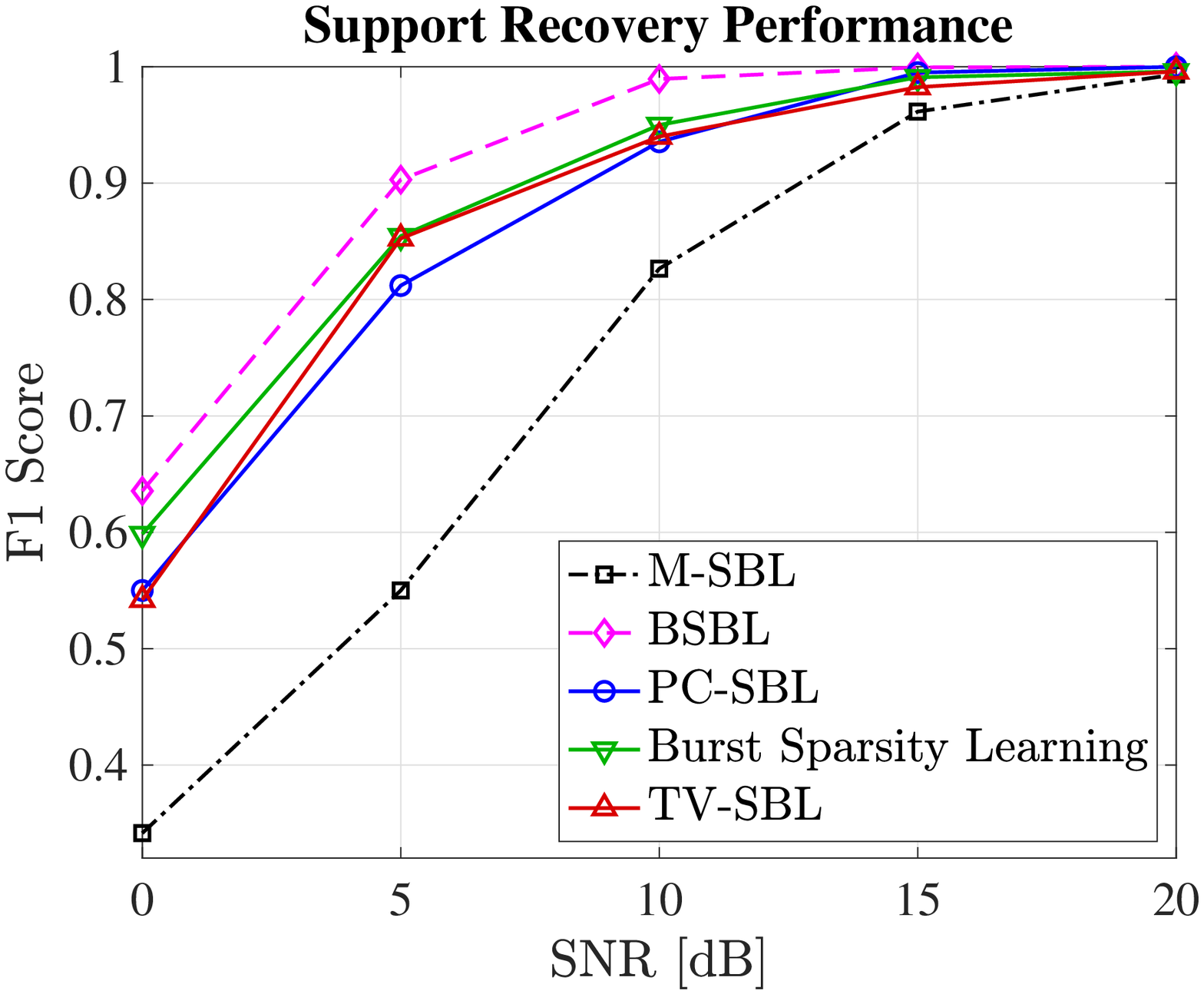}}\vspace{-4mm}
\caption{Recovery performance for homogeneous block-sparsity: (a) 2 blocks of length 5, (b) NMSE, and (c) F$_1$-Score.}\vspace{-2mm}
\label{fig_perf_comparison_random_M_20_N_150_blocks_5_2}
\end{figure*}
\begin{figure*}[t!]
\subfigure[]{\includegraphics[width=0.28\textwidth]{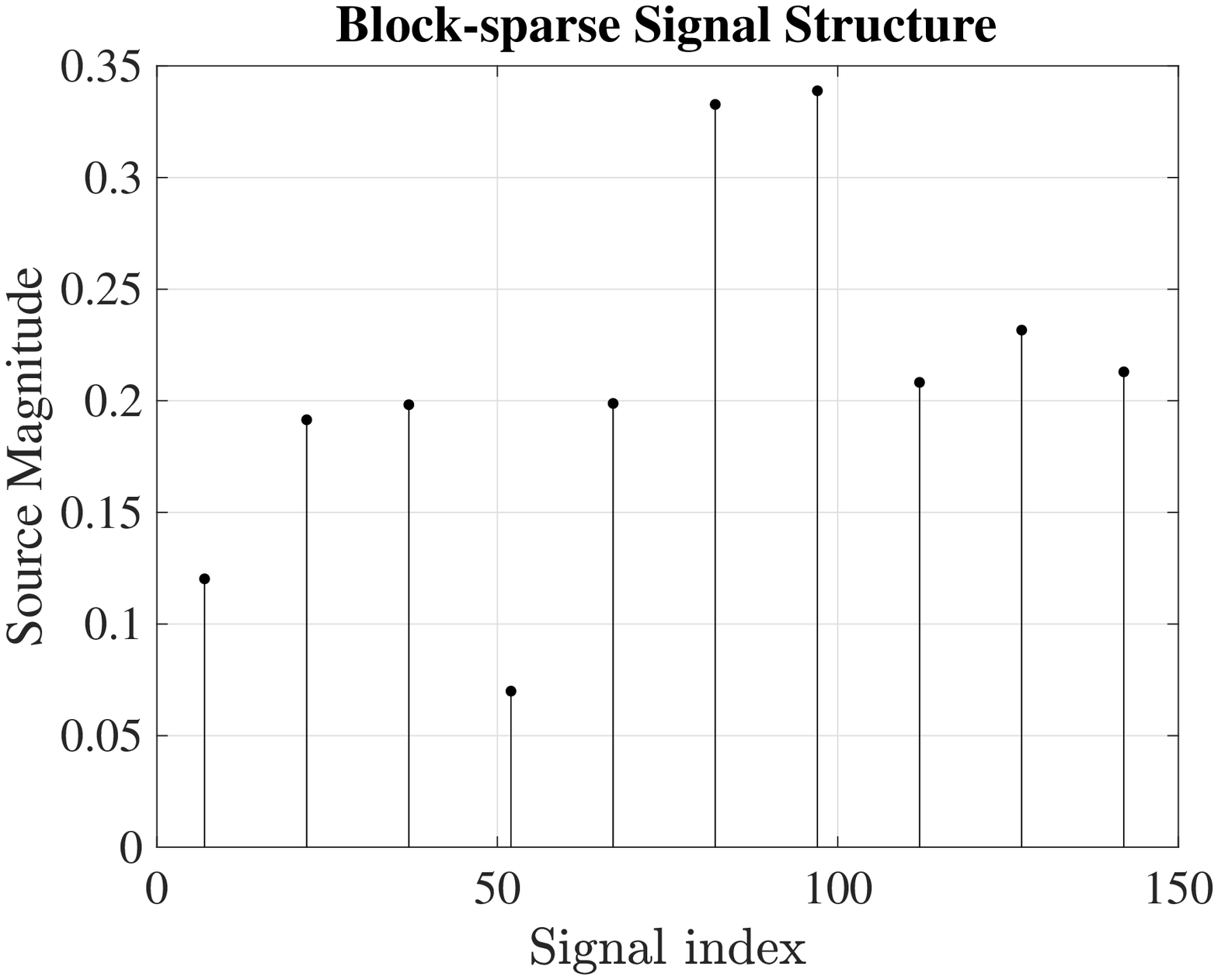}} 
\subfigure[]{\includegraphics[width=0.33\textwidth]{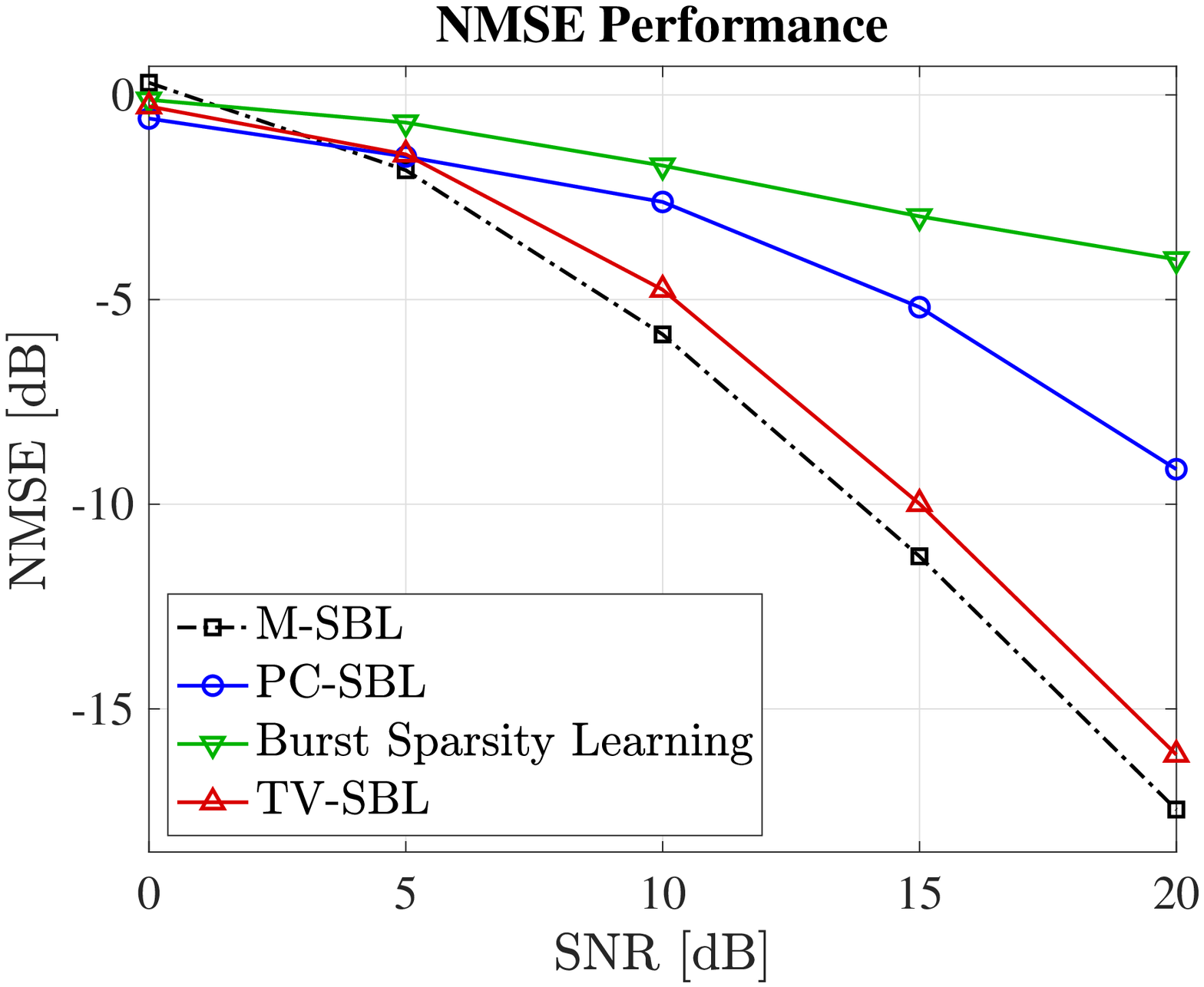}} 
\subfigure[]{\includegraphics[width=0.33\textwidth]{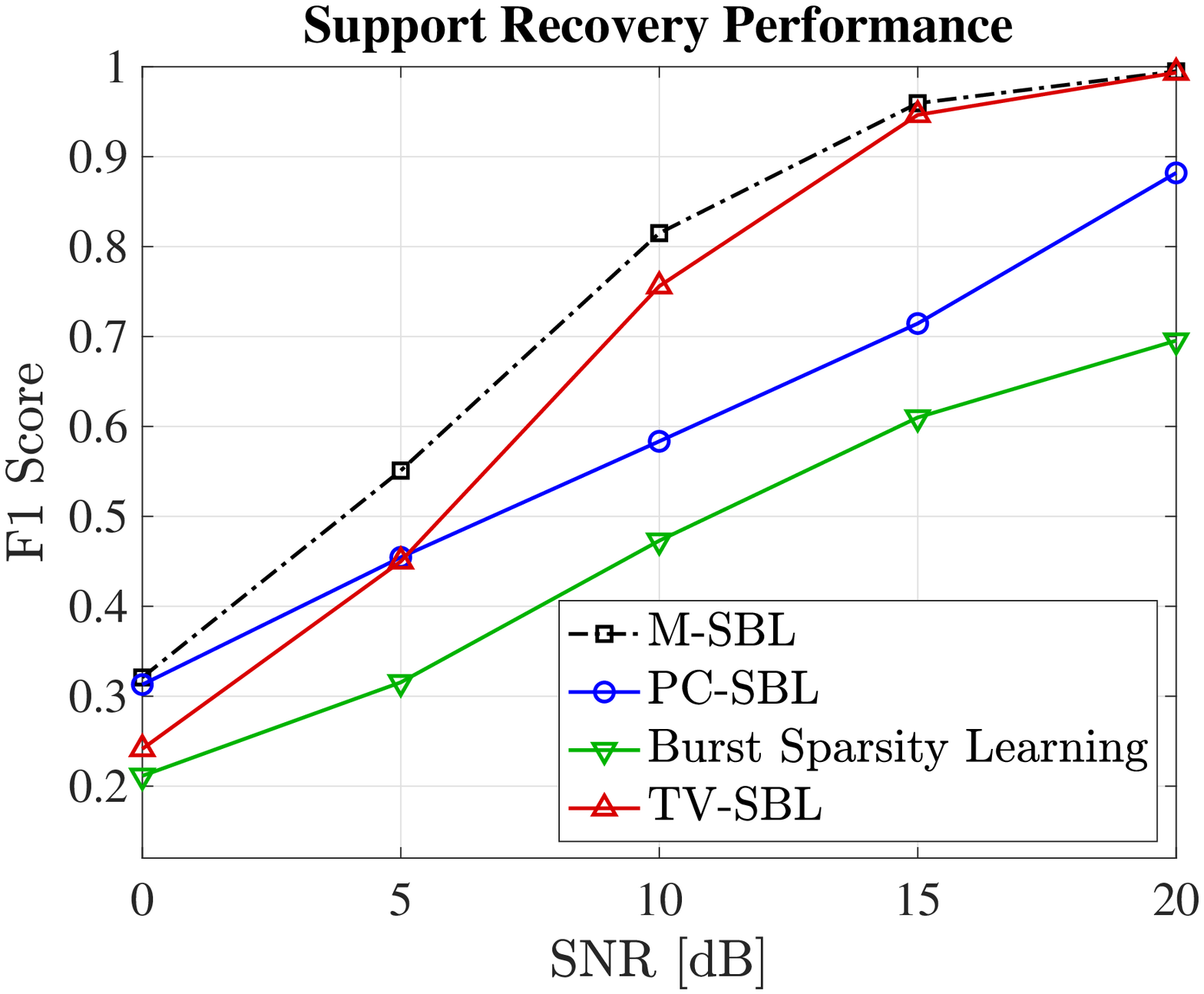}}\vspace{-4mm} 
\caption{Recovery performance for random sparsity: (a) 10 blocks of length 1, (b) NMSE, and (c) F$_1$-Score.}\vspace{-2mm}
\label{fig_perf_comparison_random_M_20_N_150_blocks_1_10}
\end{figure*}
\begin{figure*}[t!]
\subfigure[]{\includegraphics[width=0.28\textwidth]{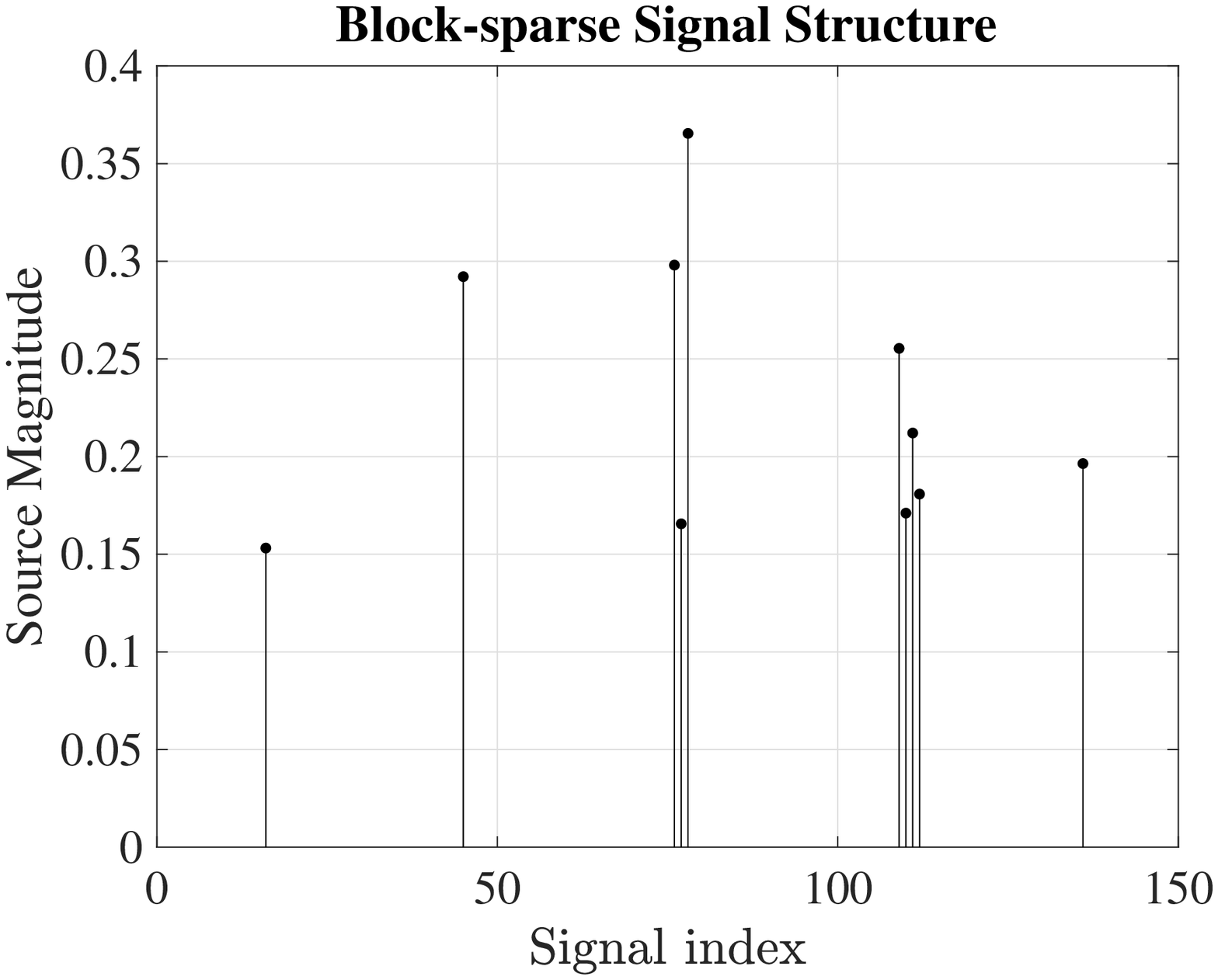}}
\subfigure[]{\includegraphics[width=0.33\textwidth]{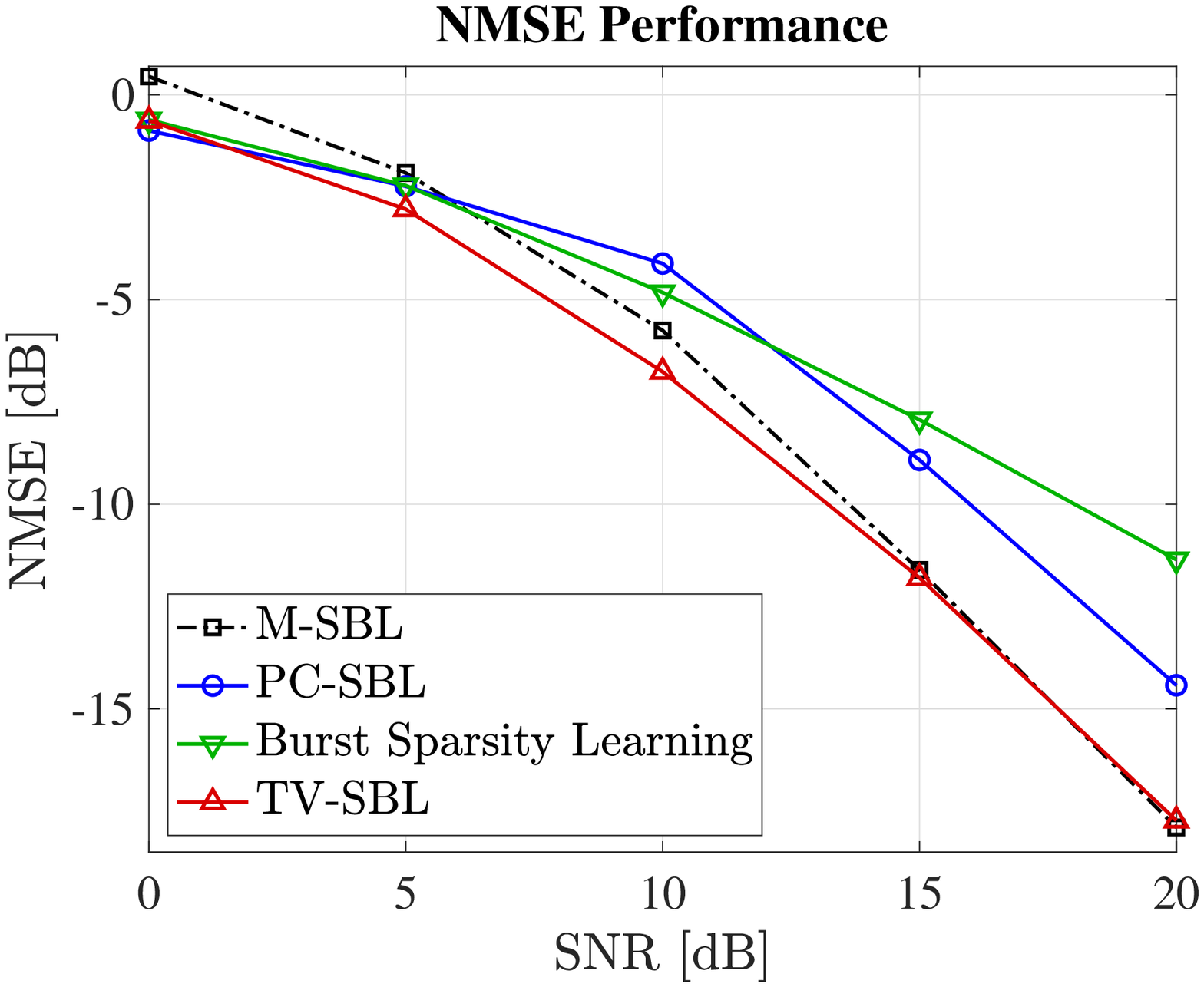}}
\subfigure[]{\includegraphics[width=0.33\textwidth]{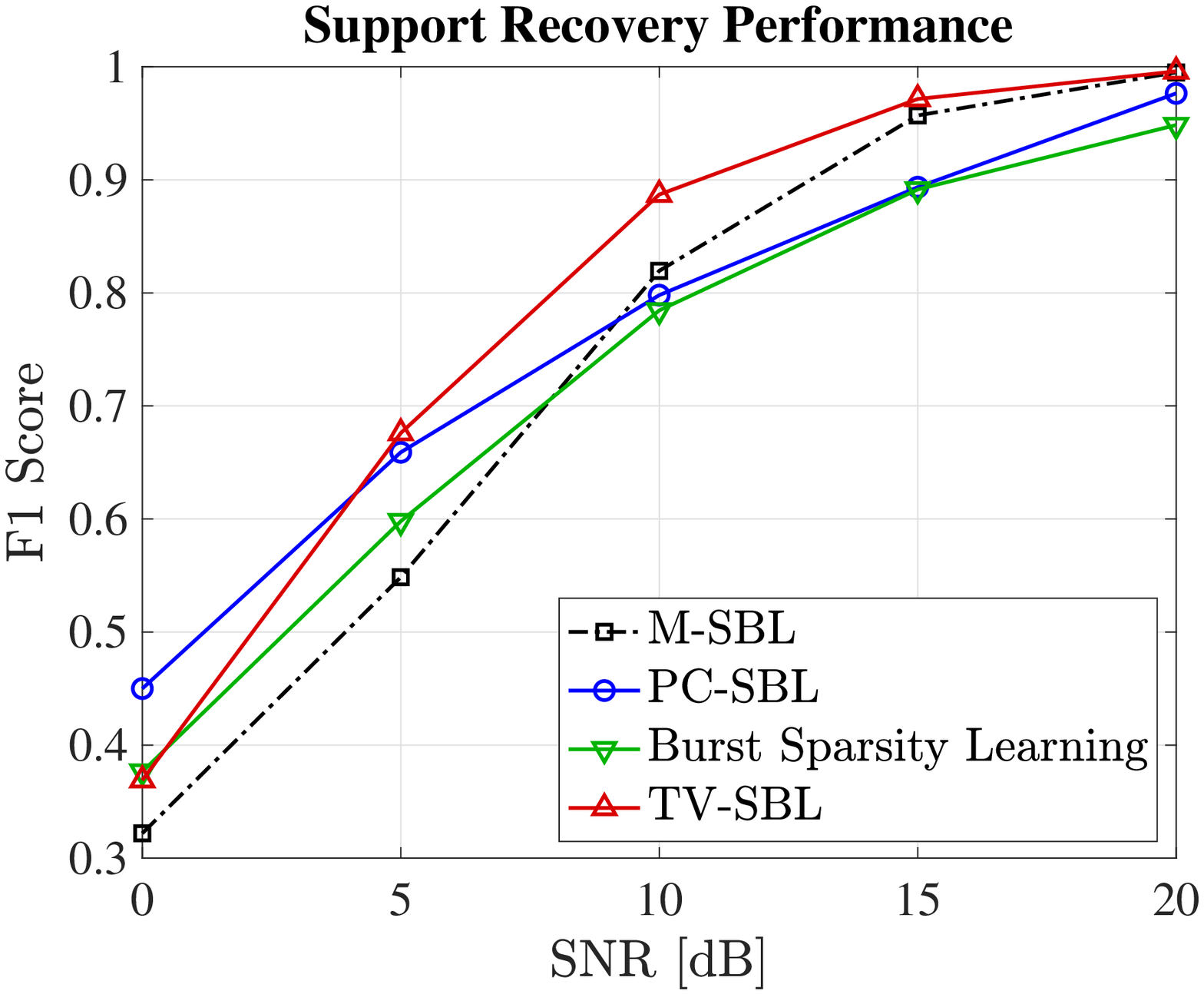}}\vspace{-4mm}
\caption{Recovery performance for hybrid sparsity: (a) 1 block of len. 4, 1 block of len. 3, and 3 blocks of len. 1, (b) NMSE, and (c) F$_1$-Score.}\vspace{-2mm}
\label{fig_perf_comparison_random_M_20_N_150_blocks_var_1}
\end{figure*}

We assess performance via the Normalized mean square error (NMSE) and support recovery. The NMSE is defined as 
$
    \mathbb{E}\left[\frac{||\hat{\Xb}-\Xb||^{2}}{||\Xb||^{2}}\right]$,
where $\hat{\Xb}$ is the estimated source matrix and the norm used is the Frobenius norm. Support recovery is evaluated using the F$_1$-Score, defined as \cite{f_score_orig}
$
    \mathrm{F}_{1} = \mathbb{E}\left[2\,\frac{\text{precision}\times\text{recall}}{\text{precision}+\text{recall}}\right]$,
where 
$\mathrm{precision} = \frac{\text{tp}}{\text{tp} + \text{fa}}$,  $ \mathrm{recall} = \frac{\text{tp}}{\text{tp} + \text{mis}}$, ``tp'': number of true positives, ``fa'': number of false alarms, and ``mis'': number of misdetections. 

\begin{remark}
For the ease of comparison, we evaluate the support recovery by preserving the $K$ largest rows of $\hat{\Xb}$ while setting the rest to zero. In practice, the support is estimated using a fixed threshold. 
\end{remark}

\vspace{0.1cm}
\noindent \textbf{Performance of Different TV Penalties}

\noindent Fig. \ref{fig_tv_compare_random_M_20_N_150_blocks_5_2} compares the performance of the Linear and Log TV regularizer in \eqref{eq_sbl_tv_case1} and \eqref{eq_sbl_tv_case2}, respectively, for the homogeneous block-sparse signal.
Both regularizers improve the performance from that of M-SBL. We observe an improved performance for the Log TV penalty, consistent with our claim in Sec. \ref{sec_novel_tv_regularizer}, showing that it is more adept at identifying block structures and denoising the zero rows of $\Xb$.

\vspace{0.1cm}
\noindent \textbf{Comparison with Benchmark Algorithms}

\noindent We study all three block-sparsity classes and compare the performance of our TV-SBL (Log TV) algorithm to SBL-based block-sparse recovery algorithms: \textit{(i)} BSBL \cite{bsbl_algorithms}, \textit{(ii)} PC-SBL \cite{pc_sbl}, and \textit{(iii)} Burst Sparsity Learning \cite{burst_sparsity_learning}. The M-SBL algorithm \cite{sbl_wipf_mmv} is used as a reference to show recovery without regularization. In order to assess the robustness of each algorithm to changes in block patterns, the parameters of each algorithm were empirically tuned for the homogeneous block-sparse signal over the SNR range, and then left unchanged for random and hybrid sparse signals.
 
\vspace{0.1cm}
\noindent \textit{1) Homogeneous block-sparse signals:}
As seen in Fig.\ \ref{fig_perf_comparison_random_M_20_N_150_blocks_5_2}, all algorithms, unsurprisingly, outperform M-SBL. BSBL, being provided block size and boundary information \textit{apriori}, attains the best F$_{1}$-Score (Fig.\ \ref{fig_perf_comparison_random_M_20_N_150_blocks_5_2}(c)).
Even without partition knowledge, the regularized SBL algorithms fare comparably in F$_1$-Score and even exceed BSBL in NMSE.
Only Burst Sparsity Learning, with its optimal coupling-based inference, exceeds TV-SBL which uses a softer prior. 
This illustrates that \textit{pure} block-sparse recovery requires explicit coupling of parameters for the best performance.

We now demonstrate that a softer prior for block-sparsity in TV-SBL gains in increased flexibility to block structure. 

\vspace{0.1cm}
\noindent \textit{2) Sparse signals:}
Fig.\ \ref{fig_perf_comparison_random_M_20_N_150_blocks_1_10}(a) represents the extreme scenario for the block-sparse algorithms, i.e., the block size is one. 
TV-SBL outperforms the coupling-based algorithms, being comparable to M-SBL.
Explicit hyperparameter coupling biases the algorithms to block structures, and thus renders them ineffective for isolated sparsity.
Using a softer prior, TV-SBL supports block-sparsity without such excessive bias; it is remarkably adept at isolated sparsity as well.

\vspace{0.1cm}
\noindent \textit{3) Hybrid sparse signals:}
The hybrid block structure in Fig.\ \ref{fig_perf_comparison_random_M_20_N_150_blocks_var_1}(a) is representative of a practical scenario for, e.g., MIMO wireless channel models, with varying angular spreads due to uneven scattering. 
As seen in Fig.\ \ref{fig_perf_comparison_random_M_20_N_150_blocks_var_1},
TV-SBL outperforms all other algorithms in this setting. The soft prior introduced by TV-SBL accommodates blocks as well as isolated components. To summarize, TV-SBL shows itself to be a robust block-sparse recovery algorithm.  

\section{Conclusion}
We proposed a TV-regularized SBL method for recovering signal blocks of unknown sizes and boundaries from compressive measurements. As a fresh idea, the method imposes a soft TV prior on the SBL hyperparameters to encourage block-sparse solutions. The developed iterative majorization-minimization algorithm necessitates only convex optimization tools to solve the problem, enabling a use of numerous efficient solvers. The numerical results showed that TV-SBL obtains superior trade-off between recovering block-sparse and random sparse signals. Such robustness has great utility in a diversity of practical sparse signal estimation scenarios.

Imposing a TV penalty on the SBL hyperparameters opens up several concave regularization penalties as well as fast numerical solvers. Our future analysis will study further the hyperparameter regularization from a more general perspective of the TV penalty.



\bibliographystyle{Bib/IEEEbib}
\mybibliography

\end{document}

%% file: Commands.tex
\theoremstyle{plain}

\theoremstyle{proposition}

\theoremstyle{definition}

\theoremstyle{remark}
\newtheorem{remark}{Remark}
 
\def\diag{{\mbox{diag}}}
\def\exp{{\mathrm{exp}}}

\def\log{{\mathrm{log}}}
\def\herm{^{\mbox{\scriptsize H}}}
\def\inv{^{-1}}

\def\trace{{\mathrm{Tr}}}
\def\tran{^{\mbox{\scriptsize T}}}

\def\argmin{{\mathrm{argmin}}}

\def\disp{\displaystyle}

\def\QCS-CE{{\mathrm{QCS}\text{-}\mathrm{CE}}}

\def\Ccal{{\mathcal{C}}}

\def\Ncal{{\mathcal{N}}}

\def\Cbb{{\mathbb{C}}}

\def\Rbb{{\mathbb{R}}}

\newcommand{\veclatin}[1]{{\bf{#1}}}
\def\zerob{\veclatin{0}}

\def\Ab{{\veclatin{A}}}

\def\Ib{{\veclatin{I}}}
\def\nb{{\veclatin{n}}}

\def\xb{{\veclatin{x}}}

\def\Xb{{\veclatin{X}}}
\def\yb{{\veclatin{y}}}

\def\Yb{{\veclatin{Y}}}

\def\Zb{{\veclatin{Z}}}

\newcommand{\vecgreek}[1]{{\boldsymbol{#1}}}

\def\gammab{{\vecgreek{\gamma}}}
\def\Gammab{{\vecgreek{\Gamma}}}

\def\mub{{\vecgreek{\mu}}}

\def\Sigmab{{\vecgreek{\Sigma}}}